\documentclass[manuscript]{aastex}
\usepackage{color}
\begin{document}
\newcommand{\changeP}[1]{{\textcolor{blue}{#1}}}
\received{}
\accepted{}
\journalid{}{}
\articleid{}{}
\paperid{}
\cpright{AAS}{}
\ccc{}
\slugcomment{Submitted \today\ to the {\it Astrophysical Journal}}

\shorttitle{Rotational quenching of CO due to H} 
\shortauthors{Yang et al.}

\title{Quantum Calculation of Inelastic CO Collisions with H: \\
I. rotational quenching of low-lying rotational levels}
\author{Benhui Yang and P. C. Stancil}
\affil{Department of Physics and Astronomy and the Center for
     Simulational Physics,\\  The University of Georgia,
      Athens, GA 30602, USA}
\author{N. Balakrishnan}
\affil{Department of Chemistry, The University of Nevada Las Vegas, 
 Las Vegas, NV 89154, USA}
\author{R. C. Forrey}
\affil{Department of Physics, Penn State University,
Berks Campus, Reading, PA 19610, USA} 
\author{J. M. Bowman}
\affil{Department of Chemistry, Emory University,
Atlanta, GA 30322, USA} 

\begin{abstract}
New quantum scattering calculations for rotational deexcitation 
transitions of CO induced by H collisions
using two CO-H potential energy surfaces (PESs) from Shepler et al. (2007) are reported.
State-to-state rate coefficients are computed for temperatures ranging
from 1 to 3000 K
for CO($v=0,j$) deexcitation from $j=1-5$ to all lower $j^\prime$ levels,
with $j$ being the rotational quantum number.
Different resonance structures in the cross sections are attributed to 
the differences in the anisotropy and the long-range van der Waals well depths  
of the two PESs.
These differences affect rate coefficients at low temperatures
and give an indication of the uncertainty of the results. 
Significant discrepancies are found between the current
rate coefficients and previous results computed using earlier potentials,
while the current results satisfy expected propensity rules.
Astrophysical applications to modeling far infrared and submillimeter observations 
are briefly discussed. 
\end{abstract}

\keywords{molecular processes ---- molecular data --- ISM: molecules}

\section {INTRODUCTION}

The current understanding of the chemical and physical conditions in   
diffuse clouds \citep{her01}, and for the majority of other low-temperature astrophysical
environments, is deduced primarily from observations of molecular hydrogen and CO,
being typically the most abundant molecules. CO 
can be detected in absorption in the ultraviolet (UV) and 
near infrared (IR) and in emission at wavelengths from the far IR (FIR) to the submillimeter (submm). 
As examples, CO  IR emission lines, due to rovibrational transitions, toward Orion Peak 1 and 2 were observed 
by the {\it Infrared Space Observatory (ISO)} with the
Short Wavelength Spectrometer (SWS) \citep{gon02}. 
The Long Wavelength Spectrometer (LWS) on {\it ISO} detected CO pure rotational lines 
in the C-rich planetary
nebulae NGC 7027 \citep{liu96,cer97}, in the C-rich objects AFGL 2688 and 
AFGL 618 \citep{jus00}, and in the reflection nebula NGC 1333 \citep{mol00}. 
Using {\it Herschel/HIFI}, \citet{buj10} observed the protoplanetary nebula CRL 618,
detecting high-$j$ transitions of CO in the FIR/submm.
{\it Herschel/HIFI} detection of high-$j$ lines was also reported by 
Yildiz et al. (2010) for CO rotational levels up to $j=$10 
towards the stars NGC 1333 IRAS 2A, IRAS 4A, and IRAS 4B.
Recently, Castro-Carrizo et al. (2012) carried out subarcsecond-resolution 
observations of CO line emission from
the early planetary nebula M 2-9, while
Lupu et al. (2012) performed CO redshift measurements 
in the H-ATLAS survey. A CO excitation analysis, considering
departures from local thermodynamic equilibrium (LTE) using the RADEX modeling
package \citep{tak07}, was also carried out by \citet{lup12} for the
4~$\rightarrow$~3 and 10~$\rightarrow$~9 pure rotational lines. Carbon monoxide was also detected in the 
brown dwarf Gliese 229B, but the predicted CO abundance obtained from 
LTE spectral modeling was $\sim10^3$ times larger than the value computed using local chemical equilibrium models 
\citep{nol97}.

In the majority of low-density environments, the  
level populations of molecules typically depart from LTE.
As a consequence, to accurately predict spectral line intensities,
rate coefficients for collisional excitation due to the impact of the
dominant species, H$_2$, H, and He, are necessary.  
For CO rotational excitation in collision with H atoms, 
one of the earliest quantum calculations  
was carried out by \citet{chu75} using semiempirical 
potentials. They presented rate coefficients for 5-150 K.
Green \& Thaddeus (1976) performed another
early quantal scattering calculation for the CO-H system over the same temperature range, but
using different semiempirical potentials. 
From their calculations,
it was found that the collisional rate coefficients for CO with H were
significantly smaller than those of CO by H$_2$, so that the contribution of CO-H collisions 
was usually neglected in later CO emission modeling.  

Later theoretical investigations on the CO-H collision system have been 
motivated by the availability of ab initio PESs with increasing
sophistication.
An early ab initio surface was obtained by \citet{bow86}, the so-called BBH PES.
BBH has a barrier, with a calculated height of 0.087 eV \citep{wan73}, to
the formation of the radical HCO. It was demonstrated from  a later scattering calculation 
by Lee \& Bowman (1987), using the BBH surface, that  for collision energies below the barrier,
a homonuclear-like behavior was observed for the 
state-to-state rotational excitation cross
sections, i.e., the cross sections displayed an even-$\Delta j$ propensity.

Keller et al. (1996) constructed another ab initio PES for the
CO-H system using the multireference
configuration-interaction (MRCI) method. They reported that this PES, called the
WKS surface, gave better
agreement with experimental spectroscopy than the BBH surface.
Using both the BBH and WKS potentials, Green et al. (1996) performed scattering 
calculations for CO rotational excitation.  
Cross sections on the two potentials revealed significant
discrepancies, particularly for the $j=$0~$\rightarrow$~1 transition.
These calculations were revisited by
\citet[][BYD]{bala02}, who verified the differences in the cross sections obtained
in the BBH and WKS surfaces observed by \citet{gre96}. BYD, however, used the
WKS PES to compute a large set of CO-H rotational excitation data.  
Subsequently, the BYD rate coefficients have been used in a
variety of astrophysical models.  For example,
\citet{lis06}, adopted the CO-H rotational excitation rate coefficients of
BYD to investigate CO rotational populations in the diffuse
interstellar medium (ISM). Based on comparisons to observations, \citet{lis06} argued that
H impact excitation makes an important contribution to the CO rotational populations and
should be included in models.  
 As argued by Lee \& Bowman (1987), an even-$\Delta j$ propensity is likely to
be valid for CO. 
However, it was later pointed out by \citet{she07} that
the excitation rate coefficients of BYD from initial $j=0$
failed to obey the homonuclear-like propensity for even-$\Delta j$ transitions, 
because the computed $j=0\rightarrow 1$ rate coefficients are larger than
the $j=0\rightarrow 2$ rate coefficients.

In order to investigate
this apparent departure from the expected behavior,
two completely independent CO-H rigid-rotor PESs were constructed
in \citet{she07} using state-of-the-art methodologies in quantum chemistry.   
One of the PESs was computed using the 
coupled-cluster singles-and-doubles-with-perturbative-triples method (CCSD(T)) together with the 
frozen-core approximation \citep{pur82}, hereafter referred as the CCSD potential. 
In the CCSD potential calculation, the doubly augmented correlation
consistent basis set of Woon \& Dunning (1994),  d-aug-cc-pVnZ, n = T, Q, and 5, 
was adopted. 
The second surface was constructed using the complete active
space self-consistent field (CASSCF) method \citep{wer85}
with the internally contracted MRCI approach \citep{wer89} 
based on aug-cc-pVQZ basis sets,
hereafter referred to as the MRCI potential. 
Quantum-mechanical close-coupling (CC) computations of the cross sections
for initial state $j=0$ were carried out by \citet{she07} on
the two PESs at kinetic energies of 400 and 800 cm$^{-1}$.
The state-to-state cross sections on both PESs showed even-$\Delta j$ propensity 
with odd-$\Delta j$ transitions being significantly suppressed.
In a subsequent investigation of the formation and excitation of CO in
diffuse clouds \citep{lis07}, the rotational excitation temperature for
the $j$=1~$\rightarrow$~0 transition  
was recomputed using the CO-H rotational excitation rate coefficients
of \citet{gre76}. 
It was also  pointed out by Liszt (2007)  that the observations are better reproduced
with these smaller CO-H rotational excitation rate coefficients. 

As discussed in \citet{she07},
the deficiency in the long-range part of the WKS PES results in the 
inaccurate rate coefficients. For odd $\Delta j$, the rate coefficients are
typically too large, and if implemented in astrophysical models, can lead to
questionable astrophysical deductions.
However, due to the lack of more accurate rate coefficients for the CO-H system, 
the BYD values are still adopted. 
Examples include a study of molecular excitation
in interstellar shock waves by Flower and Pineau des For\^{e}ts (2010), an investigation
of the molecular regions of the supernova remnants 
IC443C, W28, W44, and 3C391 and the Herbig-Hero objects HH7 and HH54 \citep{yua11}, and in radiation thermo-chemical
models of protoplanetary disks \citep{thi12}. Therefore, 
to provide accurate rate coefficients of CO rotational excitation due to impact by H atoms, 
we extend here the calculations of Shepler et al. (2007),
carrying out computations of state-to-state quenching cross sections and rate coefficients
for CO from initial states $j=1-5$ in the  ground vibrational state to all lower $j^\prime$ levels
using the potentials presented in \citet{she07}.

\section{QUANTUM MECHANICAL APPROACH}

The quantum scattering theory of a linear rigid-rotor with a structureless atom 
has been developed  by \citet{art63}.
In the current calculations, we treat CO as a rigid-rotor with  bond-length fixed at 
its equilibrium distance.
Calculations were carried out using the quantum 
CC method and the coupled-states (CS) approximation
\citep[see, for example,][]{flo07}.
The two-dimensional CO-H interaction potential is given by $V(R,\theta)$, 
where $R$ is the separation between
 the CO center of mass and the H atom,
and  $\theta$ is the angle between $\vec{R}$ and the CO molecular axis.
 The PES was expressed as
\begin{equation}
V(R,\theta)=\sum_{\lambda}^{\lambda_{\rm max}}v_{\lambda}(R)P_{\lambda}(\textrm{cos} \theta),
\end{equation}
\label{eq_vr}
where $P_{\lambda}$ are Legendre polynomials of order $\lambda$. 

For a rotational transition from an initial state  $j$ to
a final state  $j'$,
the degeneracy-averaged-and-summed integral cross section
can be given, within the CC method, as
\begin{equation}
\sigma_{j\rightarrow j'}(E_{j})
=\frac{\pi}{(2j+1)k_{j}^2}\sum_{J=0}(2J+1)\sum_{l=|J-j|}^{J+j}
\sum_{l'=|J-j'|}^{J+j'}|\delta_{jj'}\delta_{ll'}
-S_{jj'll'}^J(E_j)|^2,
\label{eq_cross}
\end{equation}
where $S_{jj'll'}^J$ is an element of the scattering matrix and $\vec{j}$ and $\vec{l}$ denote the rotational angular momentum of the CO molecule 
and the orbital angular momentum of the collision complex, respectively. 
$\vec{J}$ is the total angular momentum given by 
 $\vec{J}=\vec{l}+\vec{j}$. 
 $k_{j}=\sqrt{2\mu(E-\epsilon_{j})}/\hbar$  is the wave vector    
for the initial channel,  $E$ the total energy, $\epsilon_j$ the rotational energy of 
CO, $E_j$ the relative translational energy for the 
initial channel, and $\mu$ the reduced mass of the CO-H system.
In the CS formulation, the integral cross section is
\begin{equation}
\sigma_{j\rightarrow j'}(E_{j})
=\frac{\pi}{(2j+1)k_{j}^2}\sum_{J=0}(2J+1)\sum_{\Omega=0}^{\Omega_{\bf max}}
(2-\delta_{\Omega0})|\delta_{jj'}
-S_{jj'}^{J\Omega}(E_j)|^2,
\label{cs_cross}
\end{equation}
where $\Omega$ is the projection of the angular momentum quantum number of the diatom
along the body-fixed axis.

The nonreactive scattering code MOLSCAT \citep{molscat} was used in all our calculations.  
The modified
log-derivative Airy propagator of \citet{ale87}, with a variable step-size, was adopted to solve 
the coupled-channel equations.  
The propagation was carried out from $R=1$ $a_0$ to a maximum distance of
$R=50$ $a_0$, where $a_0$ is the atomic unit of length (the Bohr radius). 
Cross sections were calculated for collision energies between
 $10^{-5}$  and 15,000 cm$^{-1}$.
The CC calculations have been carried out for initial kinetic energies
up to 11,000 cm$^{-1}$ on the MRCI potential.
Computations with the CS approximation were performed for the MRCI, CCSD, 
and WKS potentials above 2000 cm$^{-1}$, but CC calculations were performed for the 
latter two surfaces for collision energies less than 2000 cm$^{-1}$.
The angular dependence of the
potential expressed in Eq.~(1) was expanded to $\lambda_{\rm max}=20$
with a 22-point Gauss-Legendre quadrature.
In order to ensure convergence of the state-to-state cross sections, 
at least five to ten closed channels in the basis and a sufficient
number of angular momentum partial waves were included in our calculations. 
The CO rotational constant of 1.9225 cm$^{-1}$ \citep{lov79} was used to calculate the
rotational energy levels.

\section{RESULTS AND DISCUSSION}

\subsection{State-to-state deexcitation cross sections}

\begin{figure}
\epsscale{1.0}
\plotone{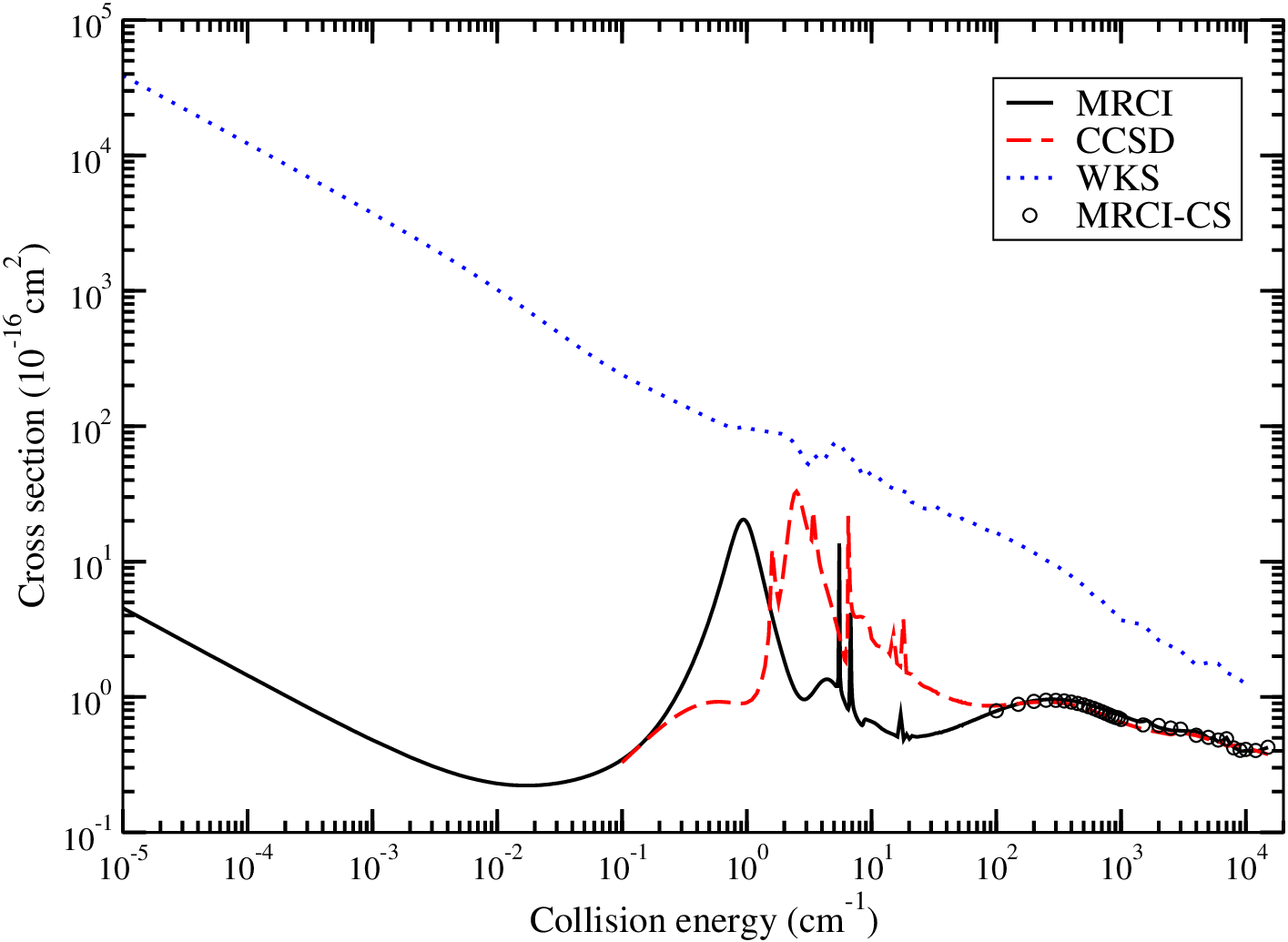}
\caption{Deexcitation cross sections of CO ($v=0$, $j$)
due to H collisions from $j=1$ to $j'=0$ as a function of kinetic energy on the 
 MRCI (solid line), CCSD (dashed line), and WKS (dotted line) potentials.
CS approximation cross sections obtained on the MRCI potential are also shown (circles),
while the CCSD potential was obtained at the CCSD(T) level of theory.
}
\label{fig1}
\end{figure}

Calculations of 
state-to-state rotational quenching cross sections have been carried out for initial 
levels $j=1, \  2, \ \cdots, \ 5$ for the deexcitation of CO($v=0,j$) by 
H.\footnote{All state-to-state 
cross sections and rate coefficients
are available on the UGA Molecular Opacity Project website
(www.physast.uga.edu/ugamop/). The rate coefficients are also available in
the format of the Leiden Atomic and Molecular Database (LAMDA) \citep{sch05} 
and in BASECOL \citep{dub06} format on our website.}
To illustrate the different behavior of the cross sections on the WKS, CCSD, and
MRCI potentials, as an example we present in Fig.~\ref{fig1}, a comparison of 
the $j=1\rightarrow 0$ deexcitation cross section.  
Due to the limited $R$ range (3-20 $a_0$) of the CCSD potential, 
the quenching cross sections are given for collision energies greater than  0.1 cm$^{-1}$.
From Fig.~\ref{fig1},  one can see that the cross sections on the WKS PES are significantly
larger than that obtained using the CCSD and MRCI potentials, 
especially at ultracold collision energies.
In the energy region between 0.1 and $\sim$20 cm$^{-1}$ , 
the cross sections computed on all three potentials exhibit 
resonances caused by the van der Waals interaction, 
but due to differences in the potential well depths \citep[see][]{she07}
very different structures are displayed, in particular the resonances on the WKS potential
are largely suppressed. 
There exists generally good agreement between the cross sections on the CCSD and MRCI PESs 
above $\sim$100 cm$^{-1}$, while the cross sections on all three surfaces appear to
converge above 10$^4$ cm$^{-1}$. 
From Fig.~\ref{fig1}, it can also be seen that the CS approximation 
is in excellent agreement with the CC method. 

\begin{figure}
\epsscale{0.9}
\plotone{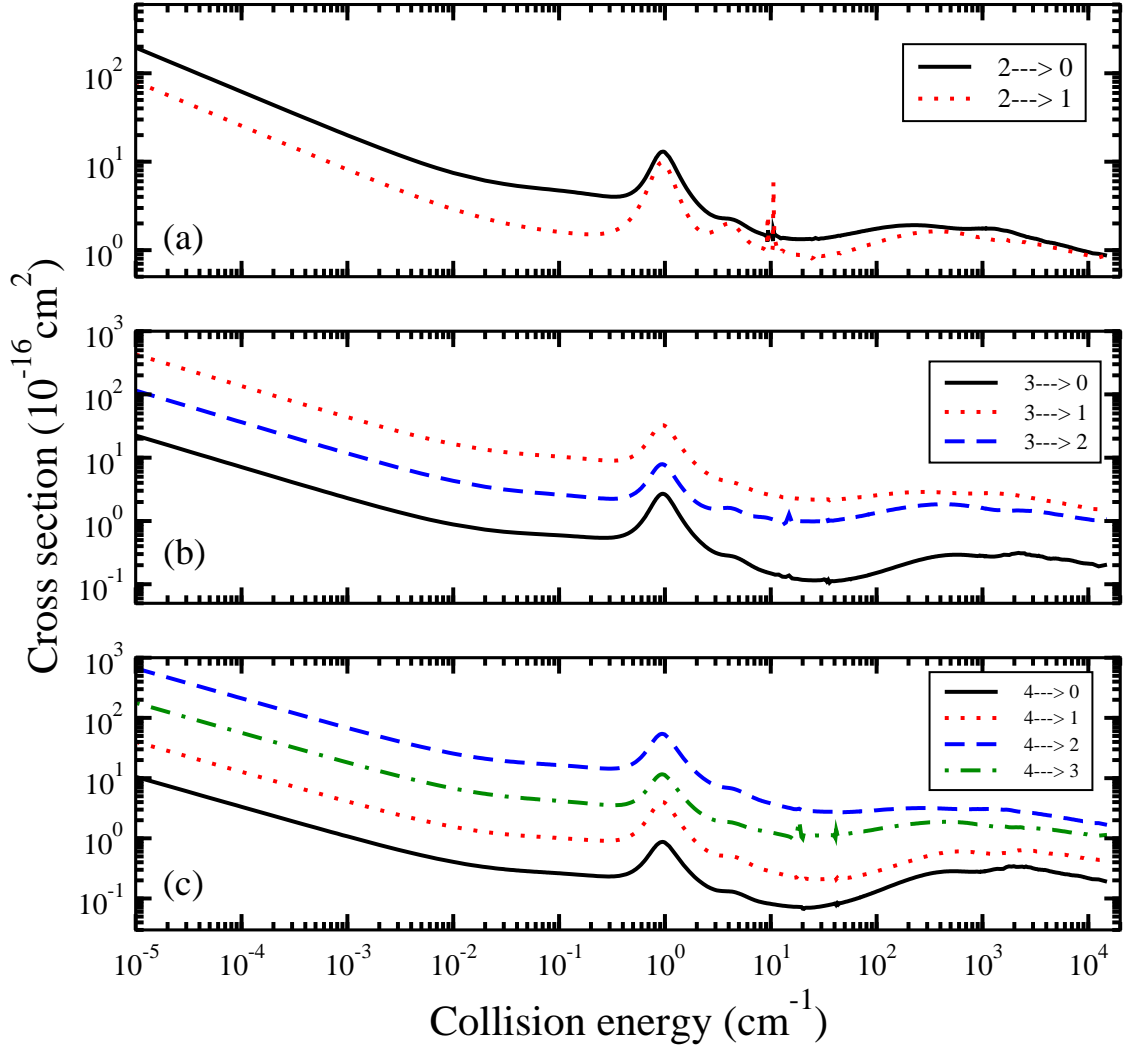}
\caption{State-to-state deexcitation cross sections of CO ($v=0$, $j$)
due to H collisions from initial states $j$=2, 3, and 4 on the MRCI potential. 
(a) $j$=2, (b) $j$=3, and (c) $j$=4. 
}
\label{fig2}
\end{figure}

For initial rotational states $j$=2, 3, 4, and 5, the  
state-to-state quenching cross sections were calculated 
using the MRCI PES of \citet{she07} only. 
Figure~\ref{fig2} presents the cross sections
from initial $j$=2, 3, and 4 into the individual final states $j'$.
Over the entire collision energy range considered, 
similar structure is observed for the different initial rotational $j$ states.  
At collision energies below $\sim$10$^{-3}$ cm$^{-1}$, only $s$-wave ($l=0$) 
scattering in the incident channel contributes, with the cross section for 
inelastic collisions varying inversely with the relative velocity,
indicative of Wigner's threshold Law  \citep{wig48}.
Similar to that shown in Fig.~\ref{fig1}, a number of resonances
at collision energies  between $\sim$1  and $\sim$30 cm$^{-1}$ are evident,
but the resonances become suppressed compared with the $j=1\rightarrow$~0 resonances
\citep[see also][]{yan06}. 
In general, the cross sections increase
with increasing $j'$, with that for $j'=0$ being the smallest for initial levels $j$=3 and 4.
For each initial $j$ state, the  $|\Delta j|=|j'-j| = 2$  transition is seen to dominate the 
quenching. 
This finding is consistent with the fact that odd-$\Delta j$ transitions in homonuclear
molecules are forbidden and since CO is nearly homonuclear, it should follow an even-$\Delta j$ propensity \citep[e.g.,][]{lee87,she07}.

\subsection{State-to-state deexcitation rate coefficients}

\begin{figure}
\epsscale{0.9}
\plotone{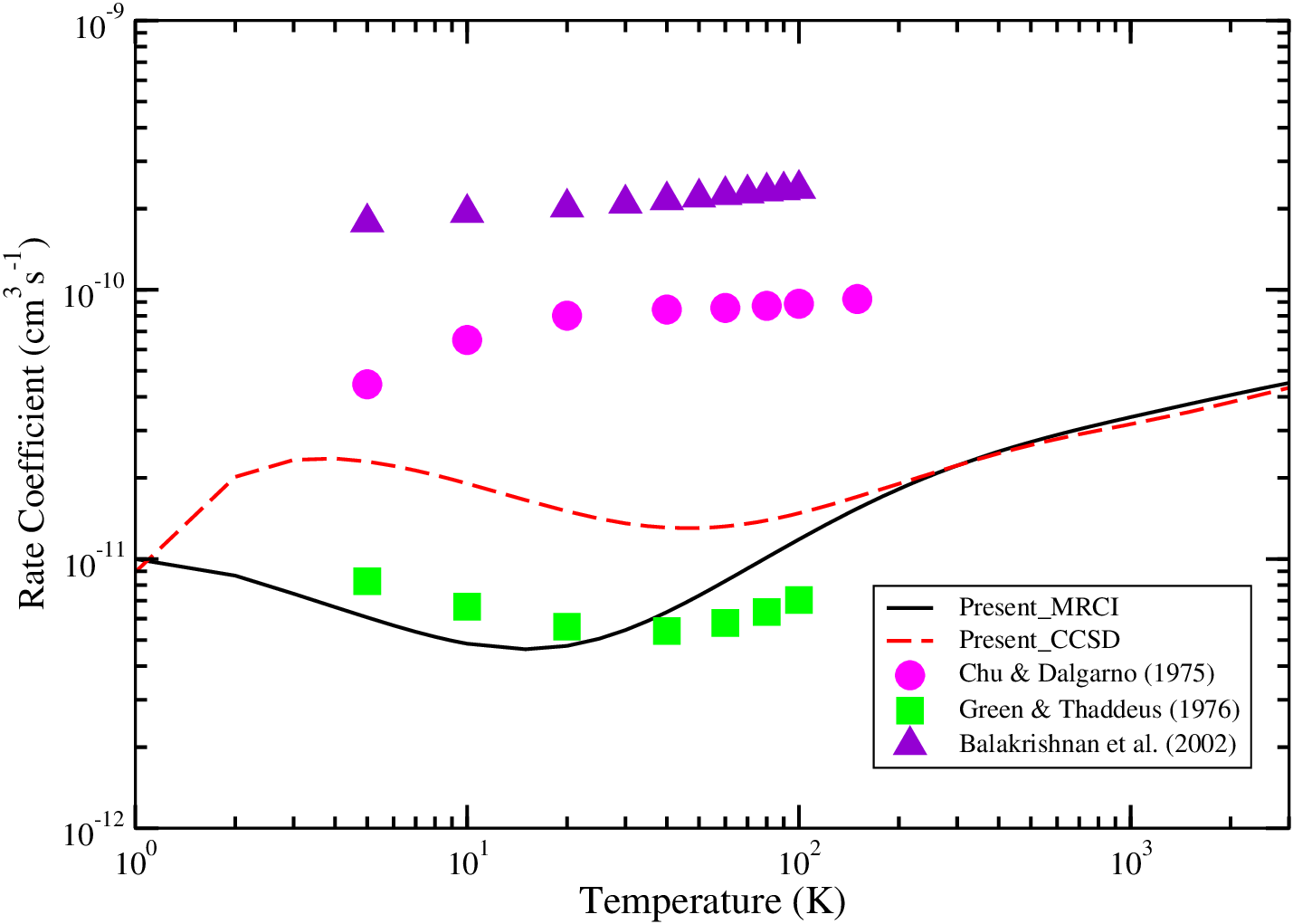}
\caption{Deexcitation rate coefficients of CO
due to H collisions from $j=1$ to $j'=0$ as a function of temperature.
The lines indicate current calculations on  
 MRCI (solid line) and CCSD (dashed line) potentials.
Symbols denote results of  
 \citet{chu75} (circle), \citet{gre76} (square), and BYD (triangle).
}
\label{fig3}
\end{figure}

By thermally averaging state-to-state quenching cross sections   
over a Maxwellian distribution of collision energy, the state-to-state quenching rate 
coefficients for deexcitation from specific initial rotational states  can be obtained 
at a temperature $T$.  For the $j=1\rightarrow j'=0$ rate coefficients
in Fig.~\ref{fig3}, we compare the present
results on the CCSD and MRCI potentials 
with previous calculations of \citet{chu75}, \citet{gre76},  and BYD.
Significant differences between the present results and those 
of \citet{chu75} and BYD are evident.   
In particular, the BYD rate coefficients obtained using the WKS potential 
are over an order of magnitude larger than the present results. 
However, the results of \citet{gre76} are in agreement with the present
rate coefficients on the MRCI PES, though their adopted semiempirical potential 
is expected to be less accurate.
As one can see from Fig.~\ref{fig3}, the current rate coefficients on the 
CCSD and MRCI potentials agree very well for temperatures above 150 K, though
differences occur for lower temperatures. The cross section resonances, as shown
in Fig.~\ref{fig1}, are responsible for the undulations in the rate coefficients.

\begin{figure}
\epsscale{0.9}
\plotone{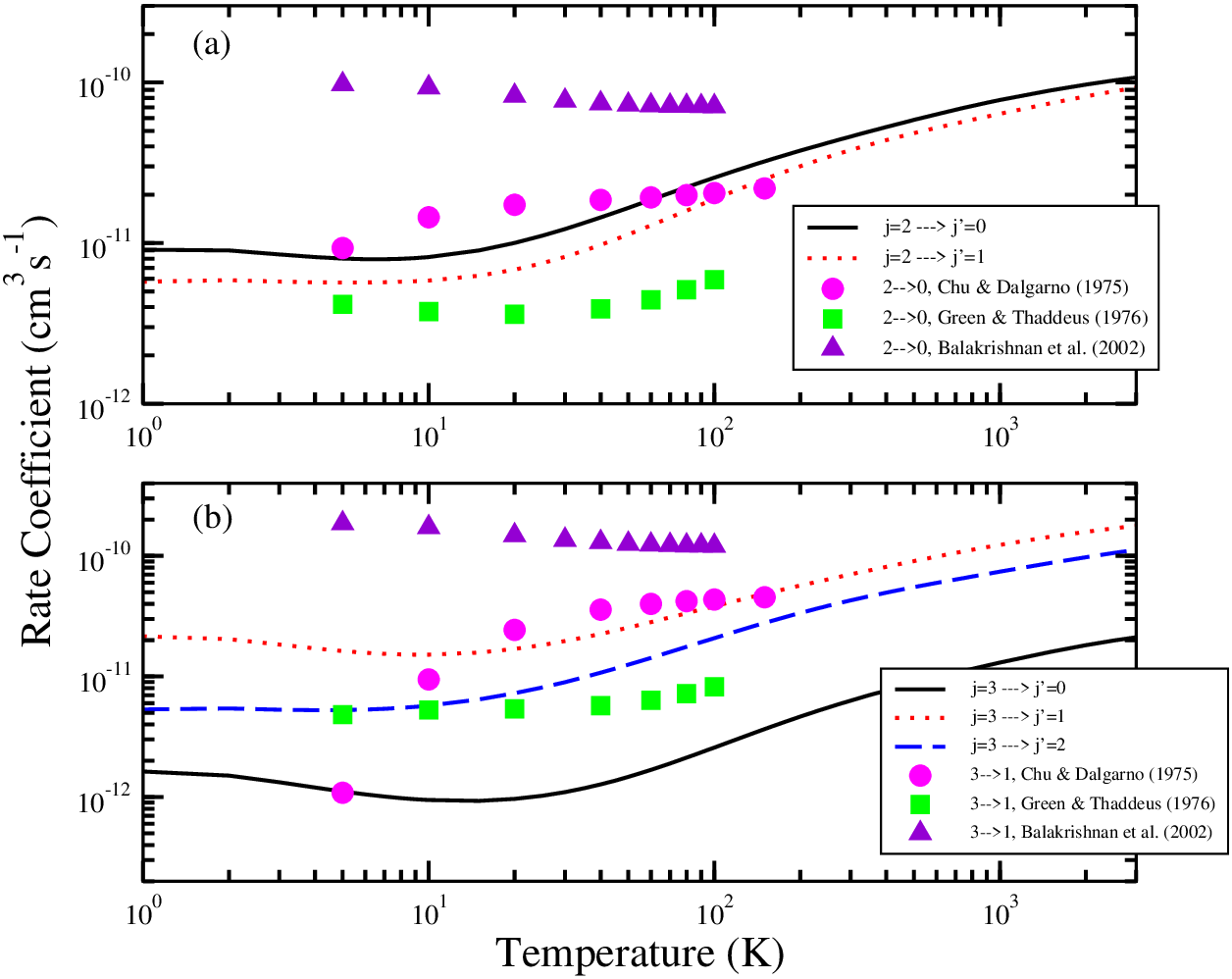}
\caption{State-to-state deexcitation rate coefficients of CO
due to H collisions from initial states $j=2$ and 3 as a function of temperature.
The lines indicate current calculations on the MRCI potential.   
Symbols denote results of   
 \citet{chu75} (circle), \citet{gre76} (square), and BYD (triangle) 
for dominant transitions. 
(a) $j=2$, dominant transition is to $j'=0$;
(b) $j=3$, dominant transition is to $j'=1$.
}
\label{fig4}
\end{figure}

\begin{figure}
\epsscale{0.9}
\plotone{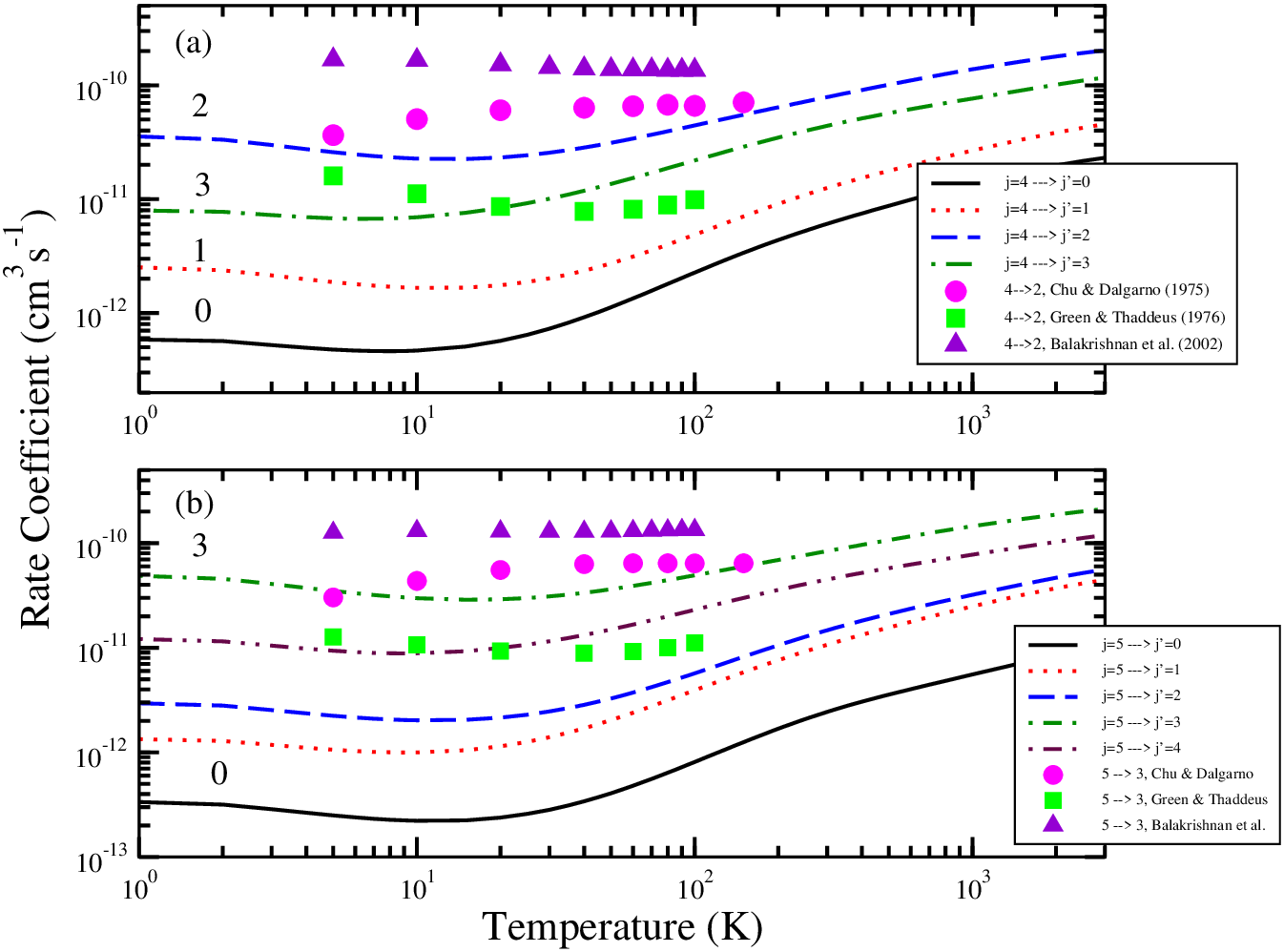}
\caption{Same as Figure \ref{fig4}, except for initial states: 
 (a) $j=4$, dominant transition is to $j'=2$;
 (b) $j=5$, dominant transition is to $j'=3$.
}
\label{fig5}
\end{figure}

For $j$=2, 3, 4, and 5,
the state-to-state deexcitation rate coefficients were computed 
using the MRCI potential only 
and are presented in Figures~\ref{fig4} and \ref{fig5} 
for temperatures ranging from 1 to 3000 K.  
For temperatures below  $\sim$20~K, the
rate coefficients display a less prominent temperature dependence  
due to suppression of the  resonances in the cross sections. 
For temperatures above $\sim$20~K, the rate coefficients show a general
increasing trend for all transitions.
Further, the rate coefficients increase with decreasing 
$\Delta j$, except for $\Delta j=-2$ transitions which dominate.
For those dominant transitions out of each initial level,
$2 \rightarrow 0$, $3 \rightarrow 1$, $4 \rightarrow 2$, and $5 \rightarrow 3$, 
 the rate coefficients
of \citet{chu75}, \citet{gre76}, and BYD are also
displayed in Figures~\ref{fig4} and \ref{fig5}. 
Similar to Fig.~\ref{fig3} for $j=1\rightarrow 0$, 
Figures~\ref{fig4} and \ref{fig5} show large differences between 
the present rate coefficients and
those of BYD as well as those of \citet{gre76}. However, 
the rate coefficients of \citet{chu75} approach the current results.

\section{ASTROPHYSICAL APPLICATIONS}

Most of the observed FIR/submm absorption and emission lines originate from rotational
transitions due to simple molecular species such as CO. 
To accurately model
molecular spectral features, in addition to spectroscopic data, a knowledge of
collisional excitation rate coefficients due to the dominant impactors is required.
As discussed in the Introduction,
carbon monoxide is nearly ubiquitous, being detected in a vast number of objects with
varying physical, chemical, and radiative properties. As a consequence, CO plays an
important role in the thermal balance and chemistry and is a key target of many modeling efforts.
For example, the radiative and collisional rates of CO were used by Martin et al. (2004) 
in estimating the gas density and kinetic temperature in a model of the Galactic Center molecular
region. In a study of the observed properties of Galactic diffuse gas, 
Pety et al. (2008) used  the $j$=1~$\rightarrow$~0 and 2~$\rightarrow$~1 rotational 
transitions of CO to perform non-LTE models of CO brightness, while
Srianand et al. (2008) detected CO absorption 
in a damped Lyman-$\alpha$ system at high redshift. 
From their modeling with the radiative transfer code RADEX \citep{tak07}, \citet{sri08} found
that the cosmic microwave background dominates the CO excitation and that 
the rotational excitation temperatures of CO are higher than measured in the  
Galactic ISM. More highly
excited CO transitions, from $j=14$ to 30, have been observed with {\it Herschel/PACS} in
the extragalactic source NGC 1068 \citep{hai12}. 

To investigate the excitation mechanism of CO rotational and rovibrational emission seen in Herbig Ae 
disks \citep[for example, the {\it Herschel/PACS} observations of][]{mee12}, 
Thi et al. (2013) have constructed a complete CO rovibrational model
which adopted existing rate coefficients of CO in collision with H, He, H$_2$, and electrons into
the radiative-photochemical code ProDiMo. Due to the incompleteness of  the
available rate coefficients,
the missing values were estimated using
various scaling laws. For the case of CO-H, the rate coefficients of BYD were used, though
they are not reliable as pointed by
Shepler et al. (2007) and indicated in Figures \ref{fig3}-\ref{fig5}. From  modeling of protoplanetary disks, \citet{thi12} found that the
abundance of atomic H
in the CO line-emitting region is significant so that CO-H collisions are important.

As CO collisional rate coefficients are needed for CO line intensity and
abundance profile modeling, the  current  CO-H rate coefficients will be very 
useful for interpreting 
observed CO emission lines. In particular, recent observations with  
{\it Herschel/PACS} and {\it Herschel/HIFI} provide a large amount of CO FIR/submm data 
for  warm gas around evolved stars. 
Modeling the CO emission with accurate collisional rate coefficients will enhance
our ability to understand the
evolution of these objects. The availability of the new rotational CO-H rate coefficients provided
here, along with our earlier CO-H$_2$ \citep{yan10} results, should improve CO modeling for the
environments discussed above and many others.  
Further, the excitation rates due to H and H$_2$ become comparable for abundance
 ratios $n_{\rm H} /n_{\rm H_2}$ of 2-3 at 10~K and 0.5-1 at 3000~K, highlighting the importance
 of H collisions in  regions with a significant atomic hydrogen fraction.

\section{CONCLUSIONS}

We have performed explicit quantum-mechanical calculations of 
rotational quenching of CO due to H collisions  
using the close-coupling approach
and the coupled-states approximation on two different interaction potential surfaces. 
Deexcitation cross sections reveal even-$\Delta j$ propensity for the CO-H system
 over the collision energy range considered as expected on physical grounds.
State-to-state quenching rate coefficients 
for initial CO rotational levels $j=$1, 2, $\cdots$, 5 to all lower $j'$ levels 
were computed for temperatures ranging from 1 to 3000 K and are available in tables formatted
for astrophysical applications. 
Discrepancies in the  cross sections and rate coefficients on the two new potentials 
are due to the differences in the well depth and anisotropy of the surfaces.
Comparison of the present rate coefficients with previous theoretical results
show significant differences primarily due to the quality of the adopted potentials. 
The present CO-H rate coefficients can be used to aid astrophysical
modeling, while quenching from higher rotational excited states as well as vibrational
transitions will be studied in the future.

\begin{acknowledgements}

 BHY, PCS, and JMB acknowledge support from NASA under Grant No.
 NNX12AF42G, while the work of NB and RCF was partially supported
by the NSF through grant PHY-1205838 and grant PHY-1203228, 
respectively.

\end{acknowledgements}

\clearpage
\newpage

\begin{thebibliography}{}

\bibitem[Ackermann et al.(2012)]{ack12} Ackermann, M., Ajello, M., Allafort, A. et al. 2012, \aap, 756, 4

\bibitem[Alexander \& Manolopoulos(1987)]{ale87} Alexander, M. H. \& Manolopoulos, D. E. 1987,
    J. Chem. Phys., 86, 2044 

\bibitem[Andersson et al.(2011)]{and11} Andersson, S., Goumans, T. P. M., \& Arnaldsson, A. 2011, 
  Chem. Phys. Lett., 513, 31 

\bibitem[Arthurs \& Dalgarno(1963)]{art63} Arthurs, A. M. \&  Dalgarno, A. 1963, Proc. Roy. Soc.,  A256, 540 

\bibitem[Balakrishnan et al.(2002)]{bala02} Balakrishnan, N., Yan, M., \&
      Dalgarno, A. 2002, ApJ, 568, 443 

\bibitem[Bowman et al.(1986)]{bow86} Bowman, J. M., Bittman, J. S., \& Harding, L. B. 1986,
      J. Chem. Phys., 85, 911 

\bibitem[Bujarrabal et al.(2010)]{buj10} Bujarrabal, V., Alcolea, J., \& Soria-Ruiz, R. 
       2010, A\&A, 521, L3 

\bibitem[Castro-Carrizo et al.(2012)]{cas12} Castro-Carrizo, A., Neri, R., Bujarrabal, V., 
  Chesneau, O., Cox, P., \& Bachiller, R. 2012, A\&A, 545, A1

\bibitem[Cernicharo et al.(1997)]{cer97} Cernicharo, J., 
       Liu, X.-W., Gonzalez-Alfonso, E. et al. 1997, \apj, 483, L65 

\bibitem[Chu \& Dalgarno(1975)]{chu75} Chu, S.-I. \& Dalgarno, A. 1975, Roy. Soc. Lond. Proc. Ser. A,
   342, 191 

\bibitem[Dubernet et al.(2006)]{dub06} Dubernet, M. L., Grosjean, A., Daniel, F. et al. 2006,
  Journal of Plasma and Fusion Research SERIES, 7, 356-357

\bibitem[Flower(2007)]{flo07}  Flower, D. R. 2007, Molecular Collisions 
in the Interstellar Medium, 2nd ed. (Cambridge Univ. Press)

\bibitem[Flower \& Pineau des For\^ets(2010)]{flo10}  Flower, D. R. \& Pineau des For\^ets, G. 2010, 
    Mon. Not. R. Astron. Soc., 406, 1745

\bibitem[Gonz\'alez-Alfonzo et al.(2002)]{gon02} Gonz\'alez-Alfonso, E., Wright, C. M., 
   Cernicharo, J. et al.  2002, A\&A, 386, 1074

\bibitem[Green \& Thaddeus(1976)]{gre76} Green, S. \& Thaddeus, P. 1976, ApJ, 205, 766 

\bibitem[Green et al.(1996)]{gre96} Green, S., Keller, H.-M., Schinke, R., \& Werner, H.-J.
      1996, J. Chem. Phys., 105, 5416

\bibitem[Hailey-Dunsheath et al.(2012)]{hai12} Hailey-Dunsheath, S., Sturm, E., Fischer, J. et al.
   2012, \apj, 755, 57

\bibitem[Herbst(2001)]{her01} Herbst, E. 2001,  Chem. Soc. Rev.,  30, 168

\bibitem[Hutson \& Green(1994)]{molscat} Hutson, J. M. \& Green, S. 1994, MOLSCAT computer code, 
     version 14, Collaborative Computational Project No. 6 of the Engineering and Physical 
     Sciences Research Council (UK)

\bibitem[Justtanont et al.(2000)]{jus00} Justtanont, K., Barlow, M. J., Tielens, A. G. G. M. et al.
    2000, A\&A, 360, 1117 

\bibitem[Keller et al.(1996)]{kel96} Keller, H.-M., Floethmann, H., Dobbyn, A. J. et al. 
    1996, J. Chem. Phys., 105, 4983


\bibitem[Lee \& Bowman (1987)]{lee87} Lee, K.-T. \& Bowman, J. M. 1987, J. Chem. Phys., 886, 215

\bibitem[Liszt(2006)]{lis06} Liszt, H. S. 2006, A\&A, 458, 507 
\bibitem[Liszt(2007)]{lis07} Liszt, H. S. 2007, A\&A, 476, 291 

\bibitem[Liu et al.(1996)]{liu96} Liu, X.-W., Barlow, M. J., Nguyen-Q-Rieu, et al.
    1996, A\&A, 315, L257 

\bibitem[Lovas et al.(1979)]{lov79} Lovas, F. J., Johnson, D. R., \& Snyder, L. E., 1979, ApJS, 41, 451 

\bibitem[Lupu et al.(2012)]{lup12} Lupu, R. E., Scott, K. S., Aguirre, J. E., et al. 2012, \apj, 757, 135

\bibitem[Martin et al.(2004)]{mar04} Martin, C. L., Walsh, W. M., Xiao, K., et al. 2004, ApJS, 150, 239

\bibitem[Meeus et al.(2012)]{mee12} Meeus, G., Montesinos, B., Mendigutia, I., et al. 2012, \aap, 544, A78 

\bibitem[Molinari et al.(2000)]{mol00} Molinari, S., Noriega-Crespo, A., Ceccarelli, C.,  et al
    2000, \apj, 538, 698 

\bibitem[Noll et al.(1997)]{nol97} Noll, K. S,  Geballe, T. R., \&  Marley, M. S. 1997,  \apj, 489, L87 

\bibitem[Pety et al.(2008)]{pet08} Pety, J., Lucas, R., \& Liszt, H. S. 2008, \aap, 489, 217

\bibitem[Purvis \& Bartlett(1982)]{pur82}Purvis, G. D. I. \& Bartlett, R. J. 1982, J. Chem. Phys., 76, 1910


\bibitem[Sch\"oier et al.(2005)]{sch05} Sch\"oier, F. L., van der Tak, F. F. S.,
   van Dishoeck, E. F., \& Black, J. H. 2005, \aap, 432, 369

\bibitem[Shepler et al.(2007)]{she07} Shepler, B. C., Yang, B. H., Dhilp Kumar, T. J., et al. 2007, A\&A, 475, L15 


\bibitem[Srianand et al.(2008)]{sri08} Srianand, R., Noterdaeme, P., Ledoux, C., \& Petitjean, P. 2008, \aap, 482, L39  

\bibitem[Thi et al.(2013)]{thi12}Thi, W. F., Kamp, I., Woitke, P., van der Plas, G., Bertelsen, 
  \& R., Wiesenfeld, L. 2013, \aap, 551, A49

\bibitem[van der Tak et al.(2007)]{tak07} van der Tak, F. F. S., Black, J. H., Sch\"{o}ier, 
 F. L., Jansen, D. J., \& van Dishoeck, E. F. 2007, A\&A, 468, 627

\bibitem[Wang et al. (1973)]{wan73} Wang, H. Y., Eyre, J. A., \& Dorfman, L. M. 1973,
   J. Chem. Phys., 59, 5199

\bibitem[Werner \& Knowles(1985)]{wer85} Werner, H.-J. \& Knowles, P. J. 1985, J. Chem. Phys., 82, 5053

\bibitem[Werner \& Knowles(1989)]{wer89} Werner, H.-J. \& Knowles, P. J. 1989, J. Chem. Phys., 89, 5803

\bibitem[Wigner(1948)]{wig48}Wigner, E. P. 1948, Phys. Rev. 73, 1002

 \bibitem[Woon \& Dunning(1994)]{woo94} Woon, D. E., \& Dunning, T. H., Jr. 1994, J. Chem. Phys., 100, 2975

\bibitem[Yang et al.(2006)]{yan06} Yang, B., Perera, H., Balakrishnan, N., Forrey, R. C., \& Stancil,
   P.  C. 2006, J. Phys. B., 39, S1229

\bibitem[Yang et al.(2010)]{yan10} Yang, B., Stancil, P. C., Balakrishnan, N., \& Forrey, R. C.
2010, \apj, 718, 1062

\bibitem[Yildiz(2010)]{yil10} Yildiz, U. A., van Dishoeck, E. F., Kristensen, L. E., et al. 2010, A\&A,, 521, L40

\bibitem[Yuan \& Neufeld(2011)]{yua11}Yuan, Y. \& Neufeld, D. A. 2011, \apj, 726, 76 


\end{thebibliography}
\end{document}